\documentclass[twocolumn,superscriptaddress,showpacs,prl,floatfix]{revtex4}
\usepackage{graphicx,epsfig,psfrag}
\usepackage{amsmath}

\newcommand{\be}{\begin{equation}}
\newcommand{\ee}{\end{equation}}
\newcommand{\bea}{\begin{eqnarray}}
\newcommand{\eea}{\end{eqnarray}}

\begin{document}
\title{
\boldmath 
Virtual Hadronic and Leptonic Contributions to Bhabha Scattering 
\unboldmath}
\author{Stefano Actis}
\affiliation{Institut f\"ur Theoretische Physik E, RWTH Aachen,      
            D-52056 Aachen, Germany}
\author{Micha{\l} Czakon}
\affiliation{Institut f\"ur Theoretische Physik und Astrophysik, Universit\"at W\"urzburg,
Am Hubland, 
D-97074 W\"urzburg, Germany}
\affiliation{Institute of Physics, University of Silesia, 
            Uniwersytecka 4, PL-40007 Katowice, Poland }
\author{Janusz Gluza}
\affiliation{Institute of Physics, University of Silesia, 
            Uniwersytecka 4, PL-40007 Katowice, Poland }
\author{Tord Riemann}
\affiliation{Deutsches Elektronen-Synchrotron, DESY, Platanenallee 6, D-15738 Zeuthen, Germany }

\begin{abstract} 
Using dispersion relations, we derive the complete virtual QED contributions
to  Bhabha scattering  due to vacuum polarization effects.
We apply our result to
hadronic corrections and to heavy lepton and top quark loop insertions. We give
the first complete estimate of their  net numerical effects for both small and
large angle scattering at typical beam energies of meson factories, LEP, and
the ILC.  
With a typical amount of 1--3 per mille they are of relevance for precision experiments.  
\end{abstract}

\maketitle 
\allowdisplaybreaks
The determination of the luminosity at lepton and hadron colliders is a
necessary task, since in many cases the normalization of the measured cross
sections is an observable of direct phenomenological interest. In practice,
this task can only be solved by selecting a particular reference process,
which is expected to generate large statistics, be as free as possible of
systematic ambiguities and predicted by the theory to suitable accuracy.  As
far as lepton colliders are concerned, the above criteria are fulfilled by
Bhabha scattering, i.e. the $e^+e^- \to e^+e^-$ process, where a precision
under the per mille level can be achieved on both the theory and the
experimental sides \cite{moenig:sfb2005,jadach:sfb2005,Montagna:ustron2007b}.

In the last few years, there has been major progress in the evaluation of the
corrections at the next-to-next-to-leading order accuracy. In fact, the
two-loop QED corrections were first evaluated in the massless case in
\cite{Bern:2000ie}.  The  photonic corrections to massive Bhabha scattering
with enhancing powers of  $\ln(s/m_e^2)$ were soon derived from that
\cite{Glover:2001ev}.  The missing constant term \cite{penin:2005kf} plus the
corrections with electron loop insertions \cite{Bonciani:2004qt,Actis:2007gi}
followed later.  Recently, the heavy fermion (or $N_f=2$) corrections
were derived in the limit $m_e^2<<m^2<<s,|t|,|u|$
\cite{Becher:2007cu,Actis:2007gi}, where $m$ is the mass of the heavy fermion,
and soon after also for arbitrary $m$, with $m_e^2<<m^2,s,|t|,|u|$
\cite{Actis:2007pn2,Bonciani:2007eh0,Actis:2007pn}.

In this letter, we present the last missing part of the virtual corrections,
the hadronic ones.

\begin{figure}[tb]
\epsfig{figure=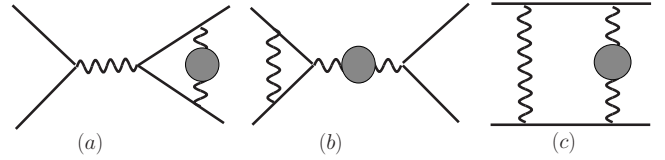,height=2.1cm}
\caption{\label{fig1}  Three classes of two-loop virtual hadronic Bhabha diagrams. (a)
and (b) represent hadronic irreducible and reducible vertex diagrams, (c)
 irreducible hadronic box diagrams.}
\end{figure}

The three classes of two-loop diagrams that we consider are shown in Fig.~\ref{fig1}.
They all may be evaluated by dispersion integrals, after replacing the vacuum
polarization insertion $\Pi_{\mathrm{had}}(q^2)$ 
to the photon propagator 
\cite{Cabibbo:1961sz},
\begin{equation}
\label{1stReplace}
\frac{g_{\mu\nu}}{q^2+i\, \delta} \, \to \,
\frac{g_{\mu\alpha}}{q^2+i\, \delta} \,
\left( q^2\, g^{\alpha\beta} - q^\alpha\, q^\beta \right) \,\Pi_{\mathrm{had}}(q^2)\,
\frac{g_{\beta\nu}}{q^2+i\, \delta},
\end{equation}
by the once-subtracted dispersion integral
\begin{equation}
\label{DispInt}
\Pi_{\mathrm{had}}(q^2) =
- \frac{q^2}{\pi} \, 
  \int_{4 M_{\pi}^2}^{\infty} \, \frac{d z}{z} \, 
  \frac{\text{Im} \, \Pi_{\mathrm{had}}(z)} {q^2-z+i\, \delta} .
\end{equation}
Finally, one relates $\text{Im} \, \Pi_{\mathrm{had}}$ to the hadronic
cross-section ratio $R_{\mathrm{had}}$,
\begin{equation}
\label{Rhad}
\text{Im} \Pi_{\mathrm{had}}(z)= 
- \frac{\alpha}{3}  R_{\mathrm{had}}(z)
= - \frac{\alpha}{3} \frac{\sigma_{e^+e^-\to \rm{hadrons}}(z)}
     {(4 \pi \alpha^2)\slash (3z)},
\end{equation}
measured experimentally  in the low-energy region and around
hadronic resonances, and given by the perturbative QCD prediction
elsewhere. For heavy fermion insertions, we have in Eq. (\ref{Rhad}) instead of $R_{\mathrm{had}}(z)$:
\begin{equation}
\label{Rf}
R_{f}(z) =
Q_f^2 C_f (1+2m_f^2/z)\sqrt{1-4m_f^2/z},
\end{equation}
with electric charge $Q_f$ and color
factor $C_f$, to leading order, which is sufficient for
practical purposes.  For leptonic or top quark intermediate states the
dispersion relation approach is just an efficient technique of evaluation, but
it becomes essential for light quark loops.  In the case of Bhabha scattering,
this method was first used some time ago for one-loop propagator insertions
\cite{Berends:1976zn}.  It was also applied to two-loop irreducible vertex
(plus soft real pair) corrections \cite{Kniehl:1988id}.  Here, we derive
semi-analytical cross section formulae for the so far unknown hadronic box
diagrams (see Fig.~\ref{fig1}c).  

After replacing the photonic self energy insertion in a two-loop diagram by
the integral (\ref{DispInt}) and  subsequently interchanging loop and
dispersion integrations, one arrives at 
   integrals given by the convolution of the hadronic data with kernel functions $K(z)$,
   where the latter represent massive one-loop Feynman diagrams.
Details of the
evaluation will be presented elsewhere.  
The one-loop self-energy kernel is trivial, $K_{\mathrm{SE}}(z)=1/(q^2-z)$.
Whereas the two-loop vertex kernel
$K_{v}$ can be found in Eq.~(5) of \cite{Kniehl:1988id}, the cross section
corrections due to the 
double boxes of Fig.~\ref{fig1}c depend on three such kernels $K_a(z),
a=A,B,C$ \cite{Actis:2007pn} \footnote{Kernels and other functions may be
downloaded at \cite{webPage:2007x3}.}.  
Notice that, unlike the vertex kernel,
the box kernels are infrared divergent, but, 
analogously to the one-loop box,
they have no singularity in the electron mass. 
The net
cross section contribution from the eight double box diagrams 
is still infrared divergent.
These boxes become, as well as the reducible diagrams with one-loop vertices and boxes, infrared finite after adding real soft photon emission.
The anatomy of that is nicely detailed in Section 2.2 of \cite{Bonciani:2004qt}.  
In order to construct an infrared-finite quantity, we combine:
(i) Born diagrams interfering with 
the two-loop box diagrams (Fig.~\ref{fig1}c) and with reducible vertices
    (Fig.~\ref{fig1}b), 
(ii) Born diagrams with a one-loop vacuum polarization function interfering with single one-loop boxes and vertices,
and finally (iii) real single-photon
    emission with a one-loop vacuum polarization \footnote{The infrared safe $N_f=2$
 {\em irreducible} vertices (see Fig.~\ref{fig1}a) and {\em pure} self energy diagrams are not included here.}.
The
resulting cross-section  becomes: 
\begin{eqnarray}
\label{sig1} 
&&\frac{ d\overline{\sigma} }{ d\Omega }=
c \int_{4M_{\pi}^2}^{\infty} dz \frac{R_{\mathrm{had}}(z)}{z}
\frac{1}{t-z} F_1(z) \\
&+&
c \int_{4M_{\pi}^2}^{\infty} \frac{dz}{z\left(s-z\right)}
\Bigl\{
R_{\mathrm{had}}(z)\Bigl[F_2(z)
 + F_3(z)\ln \bigl| 1- \frac{z}{s}\bigr| \Bigr] \nonumber \\
&-&
  R_{\mathrm{h}}  (s)\Bigl[F_2(s)+F_3(s)\ln \bigl| 1-\frac{z}{s} \bigr| \Bigr]
\Bigr\} \nonumber \\
 &+& c~\frac{R_{\mathrm{h}}  (s)}{s}\Bigl\{
F_2(s)\ln\Bigl(\frac{s}{4M_{\pi}^2}-1\Bigr)
- 6\zeta_2 F_{a}(s) \nonumber \\
&+&F_3(s)\Bigl[
2\zeta_2
+\frac{1}{2}\ln^2\Bigl(\frac{s}{4M_{\pi}^2}-1\Bigr)
+\text{Li}_2\Bigl(1-\frac{s}{4M_{\pi}^2}\Bigr)
\Bigr]
\Bigr\} , \nonumber 
\end{eqnarray}
with $c=\alpha^4/(\pi^2 s)$ and $R_{\mathrm{h}}(s) = \theta(s-4M_{\pi}^2)~R_{\mathrm{had}}(s)$.
Further,
\begin{widetext}
\begin{eqnarray}
F_1(z)&=& \frac{1}{3}\,\Bigl\{\,
9\,{\bar c}(s,t)\ln\Bigl(\frac{s}{m_e^2}\Bigr)
+\Bigl[-z^2\Bigl(\frac{1}{s}+\frac{2}{t}+2\,\frac{s}{t^2}\Bigr) 
 + z\,\Bigl( 4+4\,\frac{s}{t}+2\,\frac{t}{s}\Bigr)
+\frac{1}{2}\,\frac{t^2}{s}+6\,\frac{s^2}{t} \nonumber \\
&+& 5\,s+4\,t\Bigr] \,  \ln\Bigl(-\frac{t}{s}\Bigr)
+ s\,\Bigl(-\frac{z}{t}+\frac{3}{2}\Bigr)\,
\ln\Bigl(1+\frac{t}{s}\Bigr)
+\Bigl[\frac{1}{2}\,\frac{z^2}{s}+2\,z\,\Bigl(1+\frac{s}{t}\Bigr)-\frac{11}{4}\,s-2\,t \Bigr]\,
\ln^2\Bigl(-\frac{t}{s}\Bigr) \nonumber \\
&-&\Bigl[\frac{1}{2}\,\frac{z^2}{t} -z\,\Bigl(1+\frac{s}{t}\Bigr)+\frac{t^2}{s}
+2\,\frac{s^2}{t}+\frac{9}{2}\,s+\frac{15}{4}\,t \Bigr]\,\ln^2\Bigl(1+\frac{t}{s}\Bigr)
+\Bigl[\frac{z^2}{t}-2\,z\,\Bigl(1+\frac{s}{t}\Bigr)
+2\,\frac{s^2}{t}+5\,s+\frac{5}{2}\,t
\Bigr] \nonumber \\
&\times & \ln\Bigl(-\frac{t}{s}\Bigr)\,\ln\Bigl(1+\frac{t}{s}\Bigr) 
-4\,\Bigl[\frac{t^2}{s}+2\,\frac{s^2}{t}+3\,\Bigl(s+t\Bigr) \Bigr]\,
\Bigl[1
+\text{Li}_2\Bigl(-\frac{t}{s}\Bigr)\Bigr]
-\Bigl[ \frac{t^2}{s}+2\,\frac{s^2}{t}+3\,\Bigl(s+t\Bigr)\Bigr]\,\ln\Bigl(\frac{z}{s}\Bigr)\,
 \ln\Bigl(1+\frac{t}{s}\Bigr) \nonumber\\
&-&
\Bigl[ 2\,\frac{z^2}{t}-4\,z\,\Bigl(1+\frac{s}{t}\Bigr)
-4\,\frac{t^2}{s}
-2\,\frac{s^2}{t}+s-\frac{11}{2}\,t\Bigr]\,\zeta_2 
+\Bigl[ z^2\,\Bigl( \frac{1}{s}
+2\,\frac{s}{t^2}
+ \frac{2}{t}\Bigr)
-z\,\Bigl( \frac{t}{s}+2\frac{s}{t}+2 \Bigr)\Bigr] \ln\Bigl(\frac{z}{s}\Bigr) \nonumber \\
&-&\Bigl[ z^2\,\Bigl(\frac{1}{s}+\frac{1}{t}\Bigr) +2\,z\,\Bigl(1+\frac{s}{t}\Bigr)
+s
+2\,\frac{s^2}{t}\Bigr]\, \ln\Bigl(\frac{z}{s}\Bigr)\,
\ln\Bigl(1+\frac{z}{s}\Bigr)
+ \Bigl[\frac{z^2}{s}
+4\,z\,\Bigl(1+\frac{s}{t}\Bigr)-\frac{t^2}{s}-4\,\Bigl(s+t\Bigr)\Bigr]\,
 \nonumber \\
&\times & \ln\Bigl(\frac{z}{s}\Bigr) \ln\Bigl(1-\frac{z}{t}\Bigr)
- \Bigl[ z^2\,\Bigl(\frac{1}{s}+2\frac{s}{t^2}+ \frac{2}{t} \Bigr)
-2\, z\,\Bigl(\frac{t}{s}+2\,\frac{s}{t}
+2\Bigr)+\frac{t^2}{s}+2\,\Bigl( s+t\Bigr)\Bigr]\,
\ln\Bigl(1-\frac{z}{t}\Bigr)\nonumber\\
&+&\Bigl[ \frac{z^2}{t}-2\,z\,\Bigl(1+\frac{s}{t}\Bigr)+2\,\frac{t^2}{s}
+8\,s
+4\,\frac{s^2}{t}+7\,t \Bigr]\,
\ln\Bigl(1-\frac{z}{t}\Bigr)\,\ln\Bigl(1+\frac{t}{s}\Bigr)
+ \Bigl[\frac{z^2}{s}+4\,z\,\Bigl(1+\frac{s}{t}\Bigr)
-\frac{t^2}{s}-4\,\Bigl(s+t\Bigr)\Bigr] \nonumber \\
&\times & \text{Li}_2\,\Bigl(\frac{z}{t}\Bigr) -\Bigl[
z^2\,\Bigl(\frac{1}{s}+\frac{1}{t}\Bigr) +2\,z\,\Bigl(1+\frac{s}{t}\Bigr)
+s + 2\,\frac{s^2}{t}
\Bigr]\, \text{Li}_2\,\Bigl(-\frac{z}{s}\Bigr)
- \Bigl[\,\frac{z^2}{t}
-2\,z\,\Bigl(1+\frac{s}{t}\Bigr)
+\frac{t^2}{s}+5\,s+2\,\frac{s^2}{t}+4\,t\, \Bigr] 
\nonumber \\ 
&\times & \text{Li}_2\,\Bigl(1+\frac{z}{u}\Bigr)
\Bigr\}
+4\,{\bar c}(s,t) 
\ln\Bigl(\frac{2\,\omega}{\sqrt{s}}\Bigr)\,\Bigl[
\ln\Bigl(\frac{s}{m_e^2}\Bigr) +\ln\Bigl(-\frac{t}{s}\Bigr)
-\ln\Bigl(1+\frac{t}{s}\Bigr)-1\Bigr]
,
\\
F_2(z)&=& \frac{1}{3}\,\Bigl\{
9\,{\bar c}(t,s)
\ln\Bigl(\frac{s}{m_e^2}\Bigr)
-\Bigl[z\Bigl(\frac{t}{s}+\frac{s}{t}+2\Bigr)-5 \,
\Bigl( s+\frac{t}{2}
+\frac{1}{2}\frac{s^2}{t}\Bigr)\Bigr]\,
\ln\Bigl(-\frac{t}{s}\Bigr)
-t\Bigl( \frac{z}{s}-\frac{3}{2}\Bigr) \nonumber \\
&\times & \ln\Bigl(1+\frac{t}{s}\Bigr) +\Bigl[ \frac{z^2}{2}\,\Bigl(\frac{1}{s}+\frac{1}{t}\Bigr)
+z\,\Bigl(1+\frac{t}{s}\Bigr) +2\,\frac{t^2}{s}-\frac{s}{4}+\frac{3}{4}t
\Bigr]\,\ln^2\Bigl(-\frac{t}{s}\Bigr) \nonumber \\
&-& \Bigl[ \frac{z^2}{2\,s}-z\Bigl(1+\frac{t}{s}\Bigr)+2\,\frac{t^2}{s}
+\frac{s^2}{t}+\frac{15}{4}\,s+\frac{9}{2}\,t\Bigr]  \ln^2\Bigl(1+\frac{t}{s}\Bigr)
-\Bigl(4\,\frac{t^2}{s}+\frac{s^2}{t}+4\,s+5\,t\Bigr)\,\ln\Bigl(-\frac{t}{s}\Bigr)\,
\ln\Bigl(1+\frac{t}{s}\Bigr) \nonumber \\
&-&4\, \Bigl[\, 2\,\frac{t^2}{s}+\frac{s^2}{t}+3\,\Bigl(s+t\Bigr)\,\Bigr] \Bigl[ 1+\text{Li}_2\Bigl(-\frac{t}{s}\Bigr)\,\Bigr] 
\nonumber \\
&+&  \Bigl( 12\,\frac{t^2}{s}+3\,\frac{s^2}{t}+12\,s+15\,t\Bigr)\,\zeta_2 - \Bigl[ 2\, \frac{t^2}{s}+\frac{s^2}{t}
+3\,\Bigl(s+t\Bigr)\Bigr] \ln\Bigl(\frac{z}{s}\Bigr)\,
 \Bigl[ \ln\Bigl(1+\frac{t}{s}\Bigr) - \ln\Bigl(-\frac{t}{s}\Bigr)\Bigr] \nonumber \\
&+&\Bigl[ z^2\,\Bigl(\frac{1}{t} +\frac{2}{s}+2\,\frac{t}{s^2}\Bigr)
-z\,\Bigl(\frac{s}{t}+2
+2\,\frac{t}{s}\Bigr)\Bigr]\,\ln\Bigl( \frac{z}{s} \Bigr)
-\Bigl[ \frac{z^2}{t}+4\,z\,\Bigl(1+\frac{t}{s}\Bigr)-\frac{s^2}{t}-4\,\Bigl(s+t\Bigr)\Bigr]\,
\text{Li}_2\,\Bigl(1-\frac{z}{s}\Bigr) \nonumber \\
&+&\Bigl[ z^2\Bigl(\frac{1}{s}+\frac{1}{t}\Bigr)
+2\,z\,\Bigl(1+\frac{t}{s}\Bigr)
+2\,\frac{t^2}{s}
+t \Bigr]\, \text{Li}_2\,\Bigl(1+\frac{z}{t}\Bigr) -\Bigl[ \frac{z^2}{s}-2\,z\,\Bigl(1+\frac{t}{s}\Bigr)+\frac{s^2}{t}
+2\,\frac{t^2}{s}+4\,s
+5\,t\Bigr] \nonumber \\
&\times & \text{Li}_2\,\Bigl(1+\frac{z}{u}\Bigr)
 \Bigr\} + 4\, {\bar c}(t,s) 
\ln\Bigl(\frac{2\,\omega}{\sqrt{s}}\Bigr)\,\Bigl[
\ln\Bigl(\frac{s}{m_e^2}\Bigr)
+\ln\Bigl(-\frac{t}{s}\Bigr)
-\ln\Bigl(1+\frac{t}{s}\Bigr)-1\Bigr]
,
\\
F_3(z)&=& \frac{1}{3}\,\Bigl\{\,
\Bigl[\, \frac{z^2}{s}-2\,z\,\Bigl(1+\frac{t}{s}\Bigr)+4\,\frac{t^2}{s}
+2\,\frac{s^2}{t}+7\,s+8\,t\Bigr]\,\ln\Bigl(1+\frac{t}{s}\Bigr)-\Bigl[ z^2\,\Bigl(\frac{1}{s}+\frac{1}{t}\Bigr) 
+2\,z\,\Bigl(1+\frac{t}{s}\Bigr)
+4\,\frac{t^2}{s} \nonumber \\
&+&\frac{s^2}{t}+3\,s+4\,t \Bigr] \ln\Bigl(-\frac{t}{s}\Bigr)
-\Bigl[ z^2\,\Bigl(\frac{1}{t}+\frac{2}{s}+2\,\frac{t}{s^2}\Bigr)-2\,z\Bigl(2+\frac{s}{t}
+2\,\frac{t}{s}\Bigr)
+\frac{s^2}{t}+2\,\Bigl(s+t\Bigr)\,
\Bigr]\,
\Bigr\} 
,\\
F_{a}(z)&=& \frac{1}{3}\,\Bigl\{\,
\Bigl[\, \frac{z^2}{s}-2\,z\,\Bigl(1+\frac{t}{s}\Bigr)+2\,\frac{t^2}{s}
+2\,\frac{s^2}{t}+\frac{11}{2}\,s+5\,t\Bigr]\,\ln\Bigl(1+\frac{t}{s}\Bigr)\nonumber\\
&-&\Bigl[ z^2\,\Bigl(\frac{1}{s}+\frac{1}{t}\Bigr) +2\,z\,\Bigl(1+\frac{t}{s}\Bigr)
+2\,\frac{t^2}{s}+\frac{3}{2}\,s+\frac{5}{2}\,t \Bigr]\,\ln\Bigl(-\frac{t}{s}\Bigr)
-\Bigl[ z^2\,\Bigl(\frac{1}{t}+\frac{2}{s}+2\,\frac{t}{s^2}\Bigr)-2\,z\Bigl(2+\frac{s}{t}
+2\,\frac{t}{s}\Bigr) \nonumber \\
&-&\frac{1}{2}\,\frac{s^2}{t}-s \,
\Bigr]\,
\Bigr\}.
\end{eqnarray}
\end{widetext}
We use the abbreviation ${\bar c}(s,t) = [(s+t)^2/s + (2s^2 + 2 st + t^2)/t]/3$.
The real soft photon emission is cut at a maximal photon energy $E_{\gamma}^{\max}=\omega$.
After combining with real hard photon emission from a Monte Carlo program, the dependence on this parameter disappears. This is simulated here, as usually, by setting formally $2\omega / \sqrt{s}=1$, which switches off the corresponding terms.

Of course, we can get, from Eqn. (\ref{sig1}), 
the $N_f=2$ contributions from heavy fermion loop insertions, with the
replacements $4 M_{\pi}^2 \to 4 m_f^2$, and $R_{\mathrm{had}}(z) \to R_f(z)$, see Eqn.~(\ref{Rf}) 
\footnote{In this case, the kernel integrations in Eqn.~(\ref{sig1}) may be easily performed explicitely, leading to analytical expressions with Euler trilogarithms or simpler functions.
}.    

\begin{figure}[bthp]
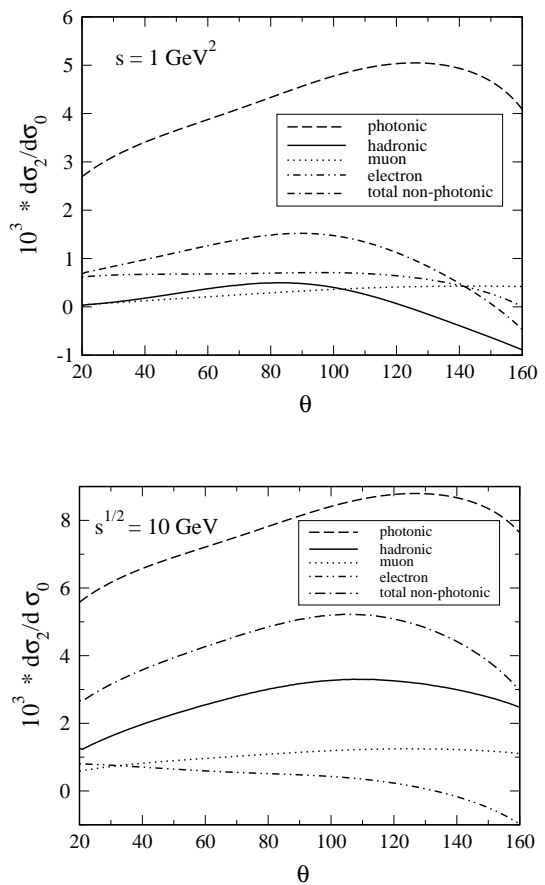

\includegraphics[scale=0.3]{PRL1gev_noSE.eps} 

\vspace*{9mm}

\includegraphics[scale=0.3]{PRL10gev_noSE.eps} 
\caption{
\label{fig-1gev}
Two-loop $N_f=2$ vertex and box corrections $d\sigma_2$ to Bhabha scattering in units of $10^{-3}d\sigma_0$ at  meson factories, $\sqrt{s}=1$ GeV (a) and $\sqrt{s}=10$ GeV (b).
}
\end{figure}

\begin{figure}[bthp]
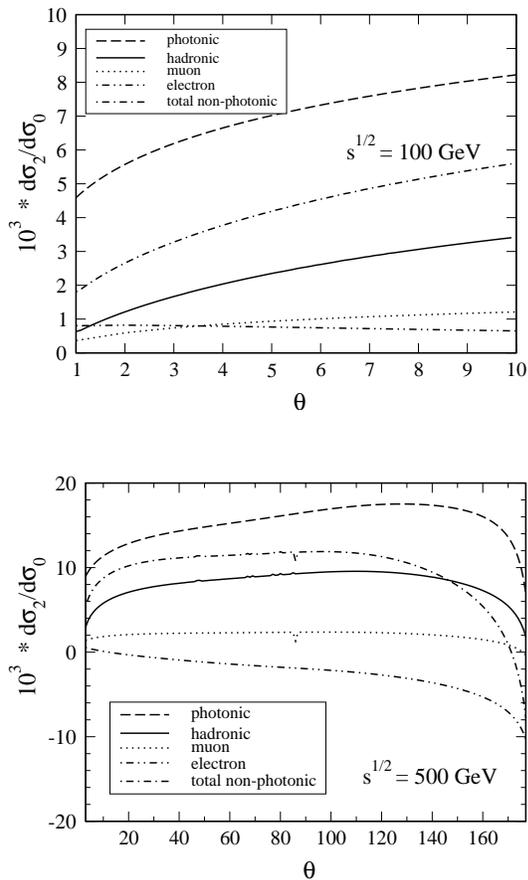

\includegraphics[scale=0.3]{PRL100gev_noSE.eps} 

\vspace*{8mm}

\includegraphics[scale=0.3]{PRL500gev_noSE.eps} 
\caption{
\label{fig-100gev}
Same as Fig.~\ref{fig-1gev}, but ILC energies of $\sqrt{s}=100$ GeV (GigaZ option) and $\sqrt{s}=500$ GeV.}
\end{figure}

We will now discuss the numerical net effects arising from the $N_f=2$ vertex plus box diagrams (i.e. excluding the pure running coupling effects):
\begin{eqnarray}
\label{dsig2}
\frac{ d\sigma_2}{ d\Omega } = \frac{ d\overline{\sigma} }{ d\Omega } + \frac{ d{\sigma}_v }{ d\Omega },
\end{eqnarray}
with $d\overline{\sigma} / d\Omega $ from Eqn.~(\ref{sig1}).
The expression for the irreducible vertex term $d{\sigma}_v / d\Omega$ derives directly from \cite{Kniehl:1988id,webPage:2007x3}.
The  $d\sigma_2/{ d\Omega }$ is normalized to the pure photonic Bhabha Born cross section $d\sigma_0/{ d\Omega }$:
\begin{eqnarray}
 \frac{d\sigma_0}{ d\Omega }  = \frac{\alpha^2}{s}\left(\frac{s}{t}+1+\frac{t}{s} 	\right)^2.
\end{eqnarray}
One has to compare the ${d\sigma_2}/{ d\Omega }$ with the anticipated
experimental accuracies varying from few tenths of per mille (at small angles
e.g. at LEP or ILC/GigaZ) to few per mille at large angles (at meson
factories, but also at LEP and the ILC) \cite{moenig:sfb2005,jadach:sfb2005,Montagna:ustron2007b}.
It also compares to the pure photonic two-loop corrections \cite{penin:2005kf}, which amount to few per mille and are relevant at all energies.
The large angle region at $\sqrt{s}=1$ GeV is studied e.g. at KLOE at DA$\Phi$NE (Fig.~ \ref{fig-1gev}a).
Here it is also important that our formulae are valid for an arbitrary ratio $s/m_f^2$ or $s/m_{\pi}^2$, as long as $m_e^2$ is small.
The hadronic corrections are very small, the net fermionic ones get about 1 per mille.
At $\sqrt{s}=10$ GeV (Fig.~ \ref{fig-1gev}b), the hadronic corrections dominate and the net corrections amount to about 3 per mille.
We also show the small angle region at $\sqrt{s} = 100$ GeV, see Fig.~ \ref{fig-100gev}b,
which is important for the luminosity determination
at LEP and the GigaZ option of the ILC.
Largest are the hadronic terms, and the net effect amounts to more than 1 per mille.
Finally, at $\sqrt{s}=500$ GeV, the net flavor terms amount to about 8 per mille and are half of the photonic effect.
In \cite{Actis:2007pn2,Actis:2007pn}, we discussed for leptons the individual terms from self-energies, irreducible vertices, and the infrared sensitive part with boxes separately. 
For the electron and heavy lepton corrections, there is also complete agreement with the results of a non-dispersive approach \cite{Bonciani:2007eh0}.
As a further check on the consistency of the
calculation we note that the corrections show  decoupling
when the masses $m_f^2$  or $M_{\pi}^2$ become large compared to the
kinematics.
In fact, the top quark and tau lepton contributions  \cite{Actis:2007pn} decouple nearly everywhere; they are included in the total non-photonic effects shown in the figures.

Naturally, the predictions depend on $R_{\mathrm{had}}$, i.e. on experimental data. 
Although $R_{\mathrm{had}}$ is being used in many problems (like the muon anomalous magnetic moment for example), no recent fit is publicly available.
So, we were forced to use an older version of  $R_{\mathrm{had}}$, as it was used in
\cite{Kniehl:1988id}, which was available by contacting the author \cite{Burkhardt:1981jk}. 
It is expected that current hadronic data would not induce changes larger than about 10\% 
in our analysis.
Because  the size of the hadronic
effects here is small enough so that only the leading digit truly
matters, this is a by far sufficient accuracy.

{\em Summarizing}, the net virtual $N_f=2$ corrections to Bhabha scattering amount
typically to 0.1\%.  They have to be taken into account for precision studies,
and a package for the evaluation has been made public at
\cite{webPage:2007x3}. 
For the electron and heavy lepton contributions, we agree with other computations.
The newly evaluated hadronic contributions are of comparable size.  
These virtual corrections have to be combined yet with cut-dependent real fermion pair or hadron emissions.
This has to be done with MC generators.
\begin{acknowledgments}
We would like to thank B.~Kniehl and H.~Burkhardt for help concerning
$R_{\mathrm{had}}$ and A. Arbuzov, H. Czyz, S.-O. Moch, and K. M{\"o}nig for
discussions. Work supported in part by GGI and INFN in Florence, by 
SFB/TRR 9 of DFG, by the Sofja Kovalevskaja Programme of the Alexander von
Humboldt Foundation, and by
MRTN-CT-2006-035505 ``HEPTOOLS'' and MRTN-CT-2006-035482 ``FLAVIAnet''.
\end{acknowledgments}
\providecommand{\href}[2]{#2}
\end{document}